\documentclass[onecolumn,showpacs,preprintnumbers,amsmath,amssymb, aps, prb]{revtex4}
\usepackage{graphicx}
\usepackage{array}
\usepackage{amsmath}
\usepackage{multirow}

\begin{document}
\preprint{}
\title{Finite element modeling of spontaneous emission of a quantum emitter at nanoscale proximity to plasmonic waveguides}
\author{Yuntian Chen}
 \affiliation{ DTU Fotonik, Department of Photonics Engineering, \O rsteds Plads, Building 345v, DK- 2800 Kgs. Lyngby,
Denmark.}
\author{Torben Roland Nielsen}
 \affiliation{ DTU Fotonik, Department of Photonics Engineering, \O rsteds Plads, Building 345v, DK- 2800 Kgs. Lyngby,
Denmark.}
\author{Niels Gregersen}
 \affiliation{ DTU Fotonik, Department of Photonics Engineering, \O rsteds Plads, Building 345v, DK- 2800 Kgs. Lyngby,
Denmark.}
\author{Peter Lodahl }
 \affiliation{ DTU Fotonik, Department of Photonics Engineering, \O rsteds Plads, Building 345v, DK- 2800 Kgs. Lyngby,
Denmark.}
\author{Jesper M\o rk}
 \affiliation{ DTU Fotonik, Department of Photonics Engineering, \O rsteds Plads, Building 345v, DK- 2800 Kgs. Lyngby,
Denmark.}

\date{\today}

\begin{abstract}

We develop a self-consistent finite element method to study spontaneous emission at nanoscale proximity of plasmonic waveguides. In the model, it is assumed that only one guided mode is dominatingly excited by the quantum emitter. With such one dominating mode assumption, the cross section of the plasmonic waveguide can be arbitrary. We apply our numerical method to calculate the coupling of a quantum emitter to a cylindrical nanowire and a rectangular waveguide, and compare the cylindrical nanowire to previous work valid in quasistatic approximation. The fraction of the energy coupled to the plasmonic mode can be calculated exactly, which can be used to determine the single optical plasmon generation efficiency for a quantum emitter. For a gold nanowire we observe agreement with the quasistatic approximation for radii below 20 nm, but for larger radii the total decay rate is up to 10 times larger. For the rectangular waveguide we estimate an optimized value for the spontaneous emission factor $\beta$ of up to $80\%$.
\end{abstract}


\maketitle
\section{Introduction}

It has long been realized that the spontaneous emission rate  is not an intrinsic property of a quantum
emitter itself  \cite{PhysRev69681}. The general explanation is that the spontaneous emission rate depends on the transition strength between the
upper and lower level of the quantum emitter and the local density
of optical states. The local density of states measures the available
number of electromagnetic modes into which the photons can be
emitted at a specific location of the emitter, and can be
manipulated by tailoring the photonic environment of the emitter.
A number of structures such as interfaces \cite{Urbach1998, johansen073303}, cavities \cite{Gunnar1991, Gerard1998}, photonic crystals \cite{Yablonovitch1987, Lodahlnature2004} and  waveguides \cite{Kleppner1981,2008PhRvL101k3903L} have already been used to modify the spontaneous emission rate. Apart from fundamental studies, engineering the spontaneous emission rate of a quantum
emitter can lead to new possibilities to boost the efficiency of  optoelectronic devices,
i.e., single-photon sources, low threshold lasers, and LED-lightening.

As an alternative to dielectric materials, the spontaneous emission rate can be manipulated by subwavelength metallic systems, which
support surface plasmon polaritons. Surface plasmon polaritons are electromagnetic
excitations associated with charge density waves on the
surface of a conducting object.  The tight confinement of the electromagnetic field to the metal-dielectric interface due to the boundary condition constraints gives the possibility of inventing new ways to enhance light-matter interaction, such as efficient single optical plasmon generation \cite{2007Natur450402A, chang053002}, single molecule detection with surface-enhanced Raman scattering \cite{Kneipp1997, citeulike2252974}, enhanced photoluminescence from quantum wells \cite{hecker1577}, and nanoantenna modified  spontaneous emission \cite{crozier4632, 2008NaPho2RT, kuhn017402}.

Although limited by the intrinsic losses of the metals in the optical frequency range, different metallic structures have been extensively studied in the last few years due to the possibilities of integration and  miniaturization.
The dramatic enhancement of the field intensity due to the field concentration and geometric slowing down of the mode propagation  provides an excellent platform to study single photon nonlinear optics \cite{2007NatPh3807C}, and light matter interaction at the single-emitter-single-photon level.
There are also considerable interests in surface plasmons for in sub-wavelength optics \cite{citeulike3441089} and applications  in sensing, near field imaging, waveguiding and switching  below the diffraction limit \cite{zayats151114, Smolyaninov2005, Takahara97, 2006Natur440508B}. The study of plasmonic effects to enhance light-matter interaction and the preferential emission of, e.g., a quantum dot into a desired mode is currently a hot research topic. It is important for solid-state quantum information devices as well as for improving our understanding of light-matter interaction at the nanoscale. So far, there are a few theoretical papers \cite{chang035420, chang053002}on this topic, and they employ simplifying assumptions that limit their applicability for analysing realistic structures, e.g. by assuming geometrical shapes that are inconsistent with current fabrication technology and making assumptions that are only valid at some length scales. The realistic description of all competing, radiative and non-radiative, decay channels for an emitter placed in close proximity to a plasmonic waveguide is important in order to understand the physics and the fundamental limitations.

The present paper  focuses on modeling of the spontaneous emission of a quantum emitter at nanoscale proximity to the realistic plasmonic waveguides by using
a finite element method, with special emphasis on  calculating the spontaneous emission $\beta$ factor. The $\beta$ factor describes the fraction of the emitted energy that is coupled to the plasmonic mode. Subwavelength waveguiding of plasmons in
metallic structures has been studied theoretically \cite{Takahara97, FanVeronis05} and  has also been  observed in a number of recent
experiments \cite{2006Natur440508B}. Enhanced spontaneous emission  of an emitter coupled to plasmonic waveguides  has been proposed \cite{chang053002, jun153111} and experimentally  demonstrated \cite{2007Natur450402A} recently. Chang  et al. \cite{chang035420} studied the spontaneous emission of an emitter coupled to a metallic nanowire by using the quasistatic approximation.
Jun et al. \cite{jun153111} employed a FDTD numerical method to study the different spontaneous emission decay rates of an emitter coupled to a metallic slot waveguide, but with assumptions for the local  density of states of the plasmonic mode.
A self-consistent model with  rigorous treatment of all the spontaneous decay rates  involved, i.e. radiative as well as non-radiative,  has not been presented in the literature. The aim of this paper is to provide such a detailed modeling.

\begin{figure}
\includegraphics[scale=0.38]{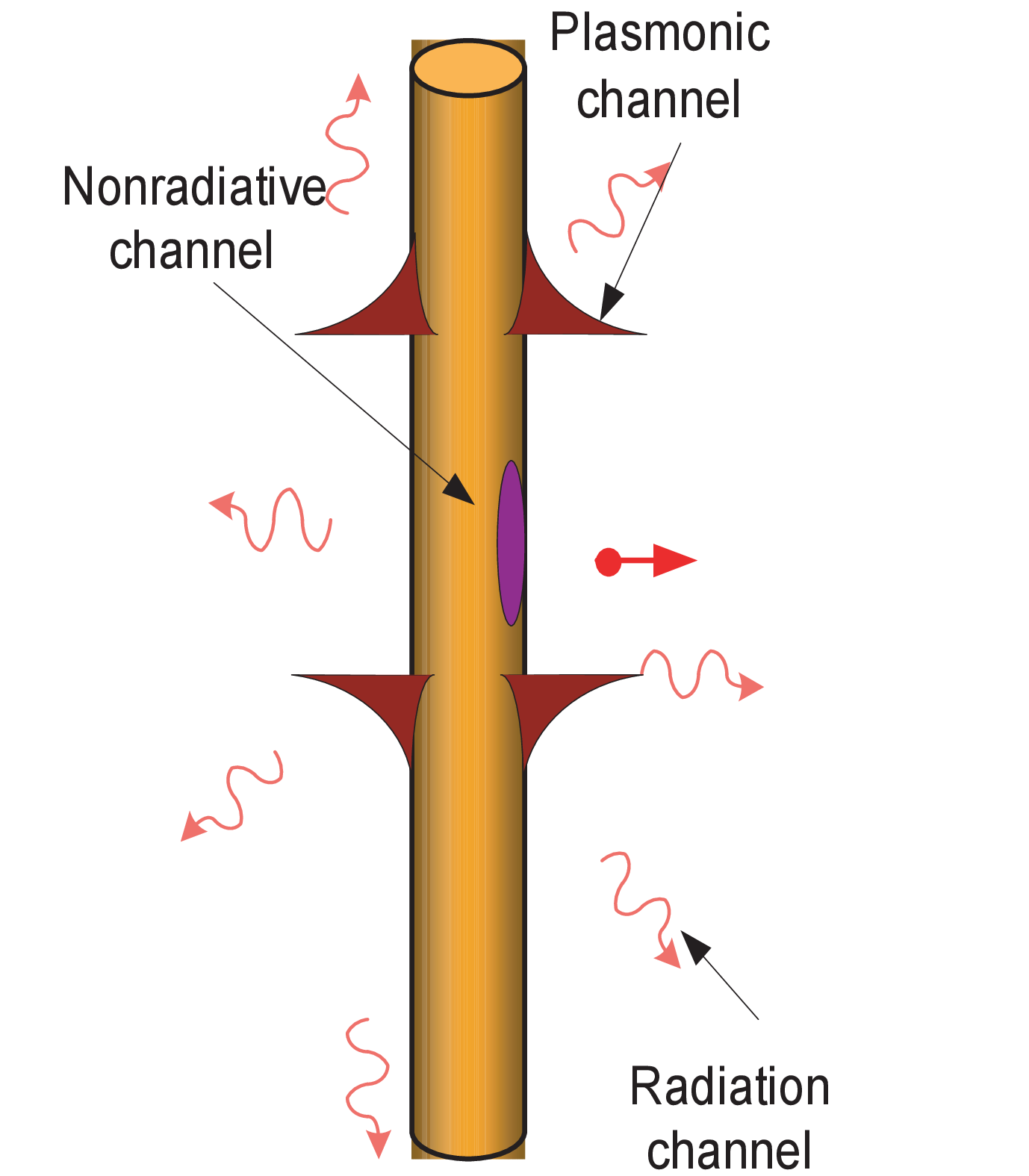}
\caption{\label{ThreeChannel}  Different emission channels involved in
the decay process of a quantum emitter (red dot) coupled to a plasmonic waveguide. In the
radiation channel the photons are traveling in  free space. In the plasmonic
channel the plasmonic modes are excited and guided by the metallic nanowire.
In the non-radiative channel,  electron hole pairs are generated.
}
\end{figure}

As shown in Fig.~\ref{ThreeChannel}, we consider an ideal quantum emitter coupled to  a plasmonic waveguide.
The excitation energy of the  quantum emitter can be
dissipated either radiatively or non-radiatively.
Radiative relaxation is associated with the emission of a photon, whereas
non-radiative relaxation can be various pathways such as coupling to
vibrations, resistive heating of the environment, or quenching by
other quantum emitters. The resistive heating of the metallic waveguide is then  the only mechanism of non-radiative relaxation considered in our model.
The quantum emitter is positioned in the vicinity of the metallic nanowire, thus there are three
channels for the quantum emitter to decay into, i.e., the radiative channel,
the plasmonic channel and the non-radiative channel. The corresponding decay rates are denoted by $\gamma_{rad}$, $\gamma_{pl}$, and $\gamma_{nonrad}$, respectively. The radiative channel is the spontaneous emission in the form of far field radiation. The plasmonic channel is
the excitation of the plasmonic mode, which is guided by the plasmonic waveguide. The non-radiative channel is
associated with the resistive heating of the lossy metals, which is due to electron-hole generation inside the metals.
The spontaneous emission $\beta$ factor is defined by  $\beta=\frac{\gamma_{pl}}{\gamma_{total}}$, where $\gamma_{total}$ is the sum of the three rates, $\gamma_{total}=\gamma_{rad}+\gamma_{nonrad}+\gamma_{pl}$. The $\beta$ factor gives the probability of the photon ``coupling'' to the plasmonic mode, when the single quantum emitter decays.


This paper is organized as follows. In Sec. II,
the computational principle and the numerical method  are presented,
First we study the dispersion relation
and the mode properties of the  plasmonic waveguide, and then
we calculate the decay rate into the plasmonic channel in a 2D model
by taking advantage of the translation
symmetry of the waveguides.  Finally, the wave  equation with a current
source in a 3D model is solved numerically, and the total decay rate of
the quantum emitter is extracted  by calculating the normalized total power
emission of the current source. Section III presents the results and discussion obtained by applying the numerical method to two different plasmonic waveguides. Section IV concludes the paper.



\section{Computational approach}

\subsection{Dispersion relation and decay rate into the plasmonic channel}

The starting point of the numerical analysis of the waveguide is the wave equation for the electric field,
\begin{displaymath}
\nabla  \times \left( {\frac{{\nabla  \times \bar E}}{{\mu _r }}} \right) - k_0^2 \varepsilon _r \bar E = 0,
\end{displaymath}
 where $k_0= \omega \sqrt{\varepsilon _0 \mu _0}$ is the vacuum wave number, $\varepsilon _r$ denotes the relative dielectric constant
 and $\mu _r$ represents the relative permeability constant, which is a constant in our model. Due to the invariance along the Z axis, the Z-dependence of the solution to the wave equation must be that of  a plane wave (complex exponential),
\begin{displaymath}
\bar E(x,y,z) = \bar E_\alpha (x,y)e^{j(\omega t -\beta z )}.
\end{displaymath}
Through out the paper, $j$ denotes $\sqrt{-1}$. For the guided plasmonic modes, at a specific frequency $\omega$ two quantization indices are needed to specify a complete set of orthogonal modes, i.e., $\alpha =\{p, \beta \}$. $\beta$ denotes the propagation constant (the component of the wave
vector along the Z-axis), and the index $p$ represents the polarization of the mode. The waveguide structure examined consists of two regions $\Omega$  and $\Lambda$.   $\Omega$ is the lossy metal core, which is surrounded by an infinite lossless dielectric medium $\Lambda$. The transverse component of the wave vector fulfills $(jk_{i \bot })^2  + \beta ^2  = \frac{{\omega ^2 }}{{c^2 }}\varepsilon _i$ with $i\in \left[\Omega,  \Lambda\right]$,
where $k_{i \bot } $   and $\varepsilon _i $   are the transverse component of the wave  vector and the relative permittivity.


%

\begin{figure}
\includegraphics[scale=0.65]{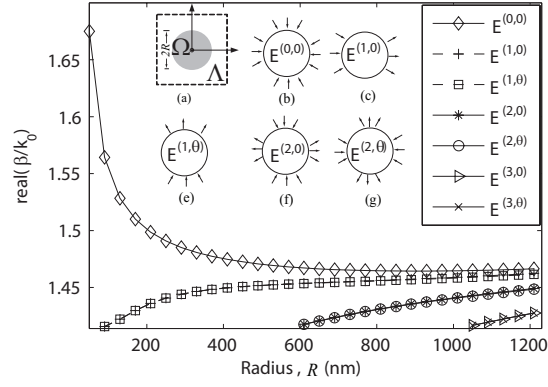}
\caption{\label{dispersion_nanowire} Dispersion relation versus radius for the metallic nanowire . Inset (a) shows the waveguide structure. Inset (b-g) show field orientation of the possible  eigenmodes supported by the waveguide.
}
\end{figure}

\begin{figure}
\includegraphics[scale=0.65]{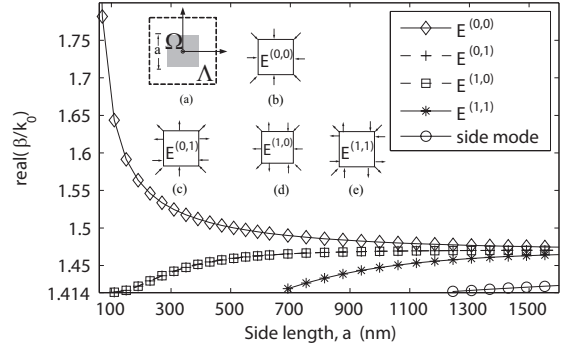}
\caption{\label{dispersion_square} Dispersion relation versus side length of the square plasmonic waveguides . Inset (a) shows the waveguide structure. Inset (b-e) shows field orientation of the possible  eigenmodes supported
by  square plasmonic waveguides.
}
\end{figure}

The finite element method can be utilized as a numerical tool to calculate the guided plasmonic modes.  The infinite dielectric medium is truncated to perform the finite  element analysis of the waveguide structure by placing the structure inside a computational window, which is large enough to guarantee the field vanishing at the boundary. Here, we consider an optical wavelength of 1 $\mu m$ and the optical permittivities of the waveguide are $\varepsilon _\Omega   = -50-3.85i$  and $\varepsilon _\Lambda   = 2$, corresponding to gold and polymer \cite{citeulike4097231}. The dispersion and the field orientation of the possible modes for cylindrical and square waveguides are presented in Fig.~\ref{dispersion_nanowire} and  Fig.~\ref{dispersion_square} respectively. As shown in the inset of Fig.~\ref{dispersion_nanowire}, these modes can be presented by two indices, where the first index denotes the number of the angular moment, $m$, and the second index describes the polarization degenerate mode with the same $m$. For example, if the $E^{m,0}$ denotes the mode with angular moment of $m$,  then $E^{m,\theta}$ denotes the corresponding degenerated mode, the field distribution of which is  rotated by $\theta$ along the $z$ axis compared with $E^{m,0}$, where  $\theta= 2\pi/2m$. As pointed out by Takahara et al. \cite{Takahara97}, the fundamental mode $E^{(0,0)}$ does not have a cutoff size of the radius, which is confirmed from the dispersion relation in Fig.~\ref{dispersion_nanowire}. The modes supported by the  metallic nanowire preserve the cylindrical symmetry of the waveguide. Due to the constraints from the boundary conditions, only TM modes exist.  For the square surface plasmon-polariton waveguides, the fundamental modes, which were studied by Jung et al.\cite{jung035434},  can be labeled in terms of two indices, which  denote the number of sign changes in the dominant component of the electric field along the x and y axes respectively. Both  plasmonic waveguides support one fundamental mode ($E^{0,0}$) without any cutoff size of the metal core, and the corresponding propagation constants increase when the size of the metal core is further shrunk, which slows down the propagating plasmonic mode. Such geometric slowing down enhances the local density of the states and the coupling efficiency to the nearby quantum emitter. In the following calculations, the size of metal core is restricted  below the cutoff size of higher order modes so that only a single mode is supported.

The electric-field dyadic Green's function for a specific guided plasmonic mode is constructed from the numerical calculation of the electric field. In the following part we will explain how to construct the electric-field dyadic Green's function for one guided plasmonic mode \cite{Thomas2001}.

The electric dyadic Green function $\bar{\bar{G }}(\bar r,\omega )$ is defined by
\begin{displaymath}
[ \nabla  \times \nabla  \times  - k_0^2 \varepsilon (\bar r)]\bar{\bar{G}}(\bar r,\omega ) = \bar{\bar{I}}\delta (\bar r - \bar r'),
\end{displaymath}
where  $\bar{\bar{I}}$ is the unit dyad. Rigorously speaking, the operator defined by $L=\left[\nabla  \times \nabla  \times  - k_0^2 \varepsilon (\bar r) \right]$
does not have a set of complete and orthogonal eigenmodes due to its non Hermitian character. Without loss of generality,
we adopt biorthogonality  in the present paper to form a complete set of ``orthogonal''  modes of the waveguides initially,
and then we will end up with an approximation from the power orthogonality for our plasmonic waveguides. Suppose that $\bar E_n$ are  a set of eigensolutions  defined by $\hat L$, the biorthogonal modes $\bar E_m^\dagger$ are defined as the eigensolutions of the adjoint operator  denoted by $\hat L^\dagger$, which is obtained from the operator $\hat L$ by replacing $\varepsilon (\bar r)$ with its complex conjugate. The  biorthogonality condition is given by
\begin{equation}\label{nor_m_g}
 \int {\varepsilon (\bar r)\bar E_n
(\bar r)[\bar E_m^\dag (\bar r)]^ *} d^3 r = \delta _{nm} N_n,
\end{equation}
with the completeness relation $\sum\limits_n {\frac{{\varepsilon (\bar r)\bar E_n (\bar r)[\bar E_n^\dag (\bar r')]^ *}}
{{N_n }}}  =\bar{\bar{I}}\delta (\bar r - \bar r')$.
From the biorthogonal completeness  relation, the dyadic Green function $\bar{\bar{G }}(\bar r,\omega )$ can be constructed  from the eigenfunction expansion as follows \cite{Thomas2001},
\begin{equation}
\begin{array}{l}
\bar{\bar{G}} (\bar r,\bar r') =
\bar{\bar{G}} _{GT} (\bar r,\bar r') +
\bar{\bar{G}} _{GL} (\bar r,\bar r') \\
  = \sum\limits_n^{} {\frac{{\bar E_n (\bar r)\cdot[\bar E_n^\dag (\bar r')]^ *}}{{N_n \lambda _n }}}  + \sum\limits_n^{} {\frac{{\nabla \phi _n (\bar r)\cdot {[\nabla \phi _n ^\dag (\bar r')} ]^ *}}{{M_n k_0^2 }}}  \\
 \end{array}
\end{equation}
where the generalized transverse part of the dyadic Green's function, $\bar{\bar{G }}_{GT}$, is
constructed from the complete set of transverse eigenfunction
$\bar E_n(\bar r)$ given by,
\begin{equation}
\begin{array}{l}
  - \nabla  \times \nabla  \times \bar E_n (\bar r) + k_0^2 \varepsilon (\bar r)\bar E_n (\bar r) = \lambda _n \varepsilon (\bar r)\bar E_n (\bar r), \\
 \nabla  \cdot [\varepsilon (\bar r)\bar E_n (\bar r)] = 0, \\
 \end{array}
\end{equation}

with the eigenvalue $\lambda_n$. The longitudinal or quasistatic part $\bar{\bar{G }}_{GL}$ is constructed from
longitudinal eigenfunction that can be found from a complete set of
scalar eigenmodes $\phi _n (\bar r)$ satisfying
 \begin{equation}
  \nabla  \cdot [\varepsilon (\bar r)\nabla \phi _n (\bar r)] = \sigma _n \phi _n (\bar r)
 \end{equation}
  with the biorthogonality relation, $ \int {\varepsilon (\bar r)\nabla \phi _n (\bar r)\cdot [\nabla \phi _n ^\dag (\bar r)]^*} d^3 r = \delta _{nm} M_n $.
Since we are studying the guided plasmonic mode, which describes the field solution in the absence of electric charge ($\nabla  \cdot [\varepsilon (\bar r)\bar E_n (\bar r)] = 0$), the  longitudinal component will vanish in the following calculations.

By applying the principle of constructing the electric-field dyadic Green's function to
the case of a plasmonic waveguide,
we find the contribution to the dyadic Green's function from the plasmonic modes as
\begin{equation}\label{green_f_spp}
\bar{\bar{G}}_{pl} (\bar r,\bar r') =\sum\limits_{p}  {
 \int\limits_{ - \infty }^{ +
\infty } {\frac{{\varepsilon _{\Lambda} \bar E_ {\alpha}(\bar r )\cdot [\bar E_ { \tilde{\alpha}}^ \dag (\bar
r')]^*}e^{-j\beta (z-z')}}{{[k_0^2\varepsilon _{\Lambda}  - (\beta ^2 - k_ {\Lambda\bot} ^2 )]N}}}
d\beta}
\end{equation}
where $\tilde{\alpha}=\{p, -\beta \}$, and the normalization factor $N$ is given by
$\delta({\beta -\beta{'}} )\delta_{p p{'}}   N =
\int {\varepsilon (\bar r)\bar E  (\bar r)[\bar E^ \dag (\bar r)]^*} d^3 r
=2 \pi \delta({\beta -\beta{'}})\delta_{p p{'} }
\int {\varepsilon (\bar r) \bar E_\alpha (\bar r)[\bar E^
\dag  _{\tilde{\alpha}^{'}}(\bar r)]^*} dxdy$,
which can be further simplified  as $N= 2\pi \int {\varepsilon (\bar r)\bar E_\alpha (\bar r)[\bar E^
\dag  _{\tilde{\alpha}^{'}}(\bar r)]^*} dxdy$, where $\alpha^{'}$ denotes $\{ p^{'}, -\beta^{'} \}$. For one plasmonic mode, the expression (5) is evaluated in
closed form by the method of contour integration as the integrand decays to zero at infinity in the upper and the lower $\beta$ plane,
\begin{equation}\label{green_f_spp_contour}
\begin{split}
\bar{\bar{G}}_{pl} (\bar r,\bar r) &= j2 \pi {\frac{{\varepsilon _2 \bar E_ {\alpha_{0}}(\bar r )
\cdot [\bar E_ { {\alpha}_{0}}^ \dag (\bar
r)]^*}}{{\frac{d(k_0^2\varepsilon _2)}{d\beta} N}}} \\
&=
{\frac{{j \pi c^2 \bar E_ {\alpha_{0}}(\bar r )
\cdot [\bar E_ { \tilde{\alpha}_{0}}^ \dag (\bar
r)]^*}}{{ \omega N v_g  }}},
\end{split}
\end{equation}
where $v_g$ is the group velocity, defined by $v_g=d \omega / d \beta$.   The corresponding density of states for one plasmonic
mode can be calculated from the  dyadic Green function according to Novotny \cite{Novotny2006}, ${\rho _\mu  (r_0 ,\omega _0 ) =
6\omega[\bar n_\mu   \cdot {\mathop{\rm Im}\nolimits}
\{ \bar {\bar{G}} (r_0 ,r_0 ,\omega _0 )\} \cdot \bar n_\mu  ]  }/(\pi c^2) $, where $n_\mu $
is the unit vector of the dipole moment. If the dipole emitter is oriented along the $X$ axis,
the density of states for the plasmonic mode is given by $\rho _{pl} (\bar r,\omega)=6|E_{\alpha,x}(\bar r)|^2/(N v_g)$.
The spontaneous emission decay rate into the plasmonic mode can be calculated by $\gamma_{pl}  = \frac{{\pi \omega _0  }}{{3\hbar \varepsilon _0 }}|\mu |^2 \rho _{pl} (\bar r,\omega)$.
Normalized by the spontaneous emission decay rate in the vacuum, the emission enhancement due to the plasmonic excitation is
\begin{equation}\label{emission_enhancement_groupV}
\frac{{\gamma _{spp} }}
{{\gamma _0 }} = \frac{{6\pi ^2 c^3 E_{\alpha _0 ,X } (\bar r)[E_{\alpha _0 ,X }^ \dag  (\bar r)]^*}}
{{\omega _0^2 N\upsilon _g }}.
\end{equation}
Eq.~(\ref{emission_enhancement_groupV}) gives a general expression of the spontaneous emission decay rate into a guided mode, supported by a lossy or lossless waveguide. In dielectric waveguides, losses are generally small, and  the biorthogonal modes $\bar E_m^\dag$ can approximately be replaced by the orthogonal mode $\bar E_m$. Such an approximation is also valid for our plasmonic waveguide, where the imaginary part of the propagation constant for the fundamental mode is around $1\%$ of the real part. According to Snyder\cite{SnyderLove1983},
the group velocity can be calculated by $v_g=\int_{A_\infty  } {(\bar E \times \bar H^ *  )
\cdot \bar zdA} /\int_{A_\infty  } {\varepsilon _0 \varepsilon (r)|\bar E(r)|^2 dA} $, where $A_\infty$
denotes integration over the transverse plane. By applying the power orthogonal approximation and the explicit form of the group velocity to Eq.~(\ref{emission_enhancement_groupV}), we can reach the following expression for the plasmonic decay rate of the fundamental mode,
\begin{equation}\label{emission_enhancement_power}
\frac{{\gamma _{pl} }}{{\gamma _0 }} =\frac{{3\pi c\varepsilon _0 E_{\alpha _0 ,X } (\bar r)E_{\alpha _0 ,X }^ *  (\bar r)}}{{k_0^2 \int_{A_\infty  } {(\bar E \times \bar H^ *  ) \cdot \bar zdA} }}.
\end{equation}


\subsection{Total decay rate}
As described in the previous subsection, the well defined field components in the transverse plane of the waveguide give the possibility of constructing the dyadic Green's function numerically. The reason is that the field is concentrated around the metallic core and is decaying to zero on the borders when the  modeling domain is reasonably large. Hence,  the perfect electric conductor  boundary condition is implemented to truncate the 2D modeling domain. However, for the radiation modes, the  field components in the transverse plane of the waveguide do not vanish no matter how large the modeling domain is. Hence, it is extremely difficult to construct the dyadic Green's function numerically for the radiation modes in a similar way as for the guided mode. Therefore, we implement a 3D  model to include the radiation modes, as well as the nonradiative contributions, by solving the wave equation with a harmonic (time dependent) source term,
\begin{equation}\label{waveguide_equation}
[\nabla  \times \frac{1}{{\mu _r }}\nabla  \times  - k_0^2 \varepsilon (\bar r)]\bar E(\bar r,\omega ) + j\omega \mu _0 \bar J(\omega ) = 0.
\end{equation}
If we introduce a test function $\bar F(\bar r,\omega )$, we can construct the functional corresponding to the wave equation in the following way\cite{Jin},
\begin{widetext}
\begin{equation}\label{weak_form}
\begin{gathered}
  L = \iiint\limits_V {[\nabla  \times \frac{1}
{{\mu _r }}\nabla  \times  - k_0^2 \varepsilon (\bar r)]\bar E(\bar r,\omega ) \cdot \bar F^ *  (\bar r,\omega )}dV + \iiint\limits_V {j\omega \mu _0 \bar J(\omega ) \cdot \bar F^ *  (\bar r,\omega )}dV \hfill \\
   = \iiint\limits_V {\frac{1}
{{\mu _r }}\nabla  \times \bar E(\bar r,\omega ) \cdot \nabla  \times \bar F^ *  (\bar r,\omega )}dV - \iiint\limits_V {k_0^2 \varepsilon (\bar r)\bar E(\bar r,\omega ) \cdot \bar F^ *  (\bar r,\omega )}dV + \iiint\limits_V {j\omega \mu _0 \bar J(\omega ) \cdot \bar F^ *  (\bar r,\omega )}dV \hfill \\
   + \mathop{{\int\!\!\!\!\!\int}\mkern-21mu \bigcirc}\limits_{\partial V}
 {\bar F^ *  (\bar r,\omega ) \cdot [\frac{1}
{{\mu _r }}\bar n \times \nabla  \times \bar E(\bar r,\omega )]} ds \hfill ,\\
\end{gathered}
\end{equation}
\end{widetext}
where ${\partial V}$ denotes the surface that encloses the volume $V$, and $\bar n$ denotes the outward unit normal vector to the surface of the modeling domain,. This is the variational formulation of the wave equation, which is required to hold for all the test functions. Eq.~(\ref{weak_form}) enables us to formulate the finite element solution for such a boundary-value problem by employing the standard finite element solution  procedures, including discretization and factorization of a sparse matrix \cite{Jin}. The boundary-value problem defined by Eq.~(\ref{weak_form}) was solved by utilizing a commercial software package, COMSOL Multiphysics \footnote{http://www.comsol.com}.
\begin{figure}
\includegraphics[scale=0.32]{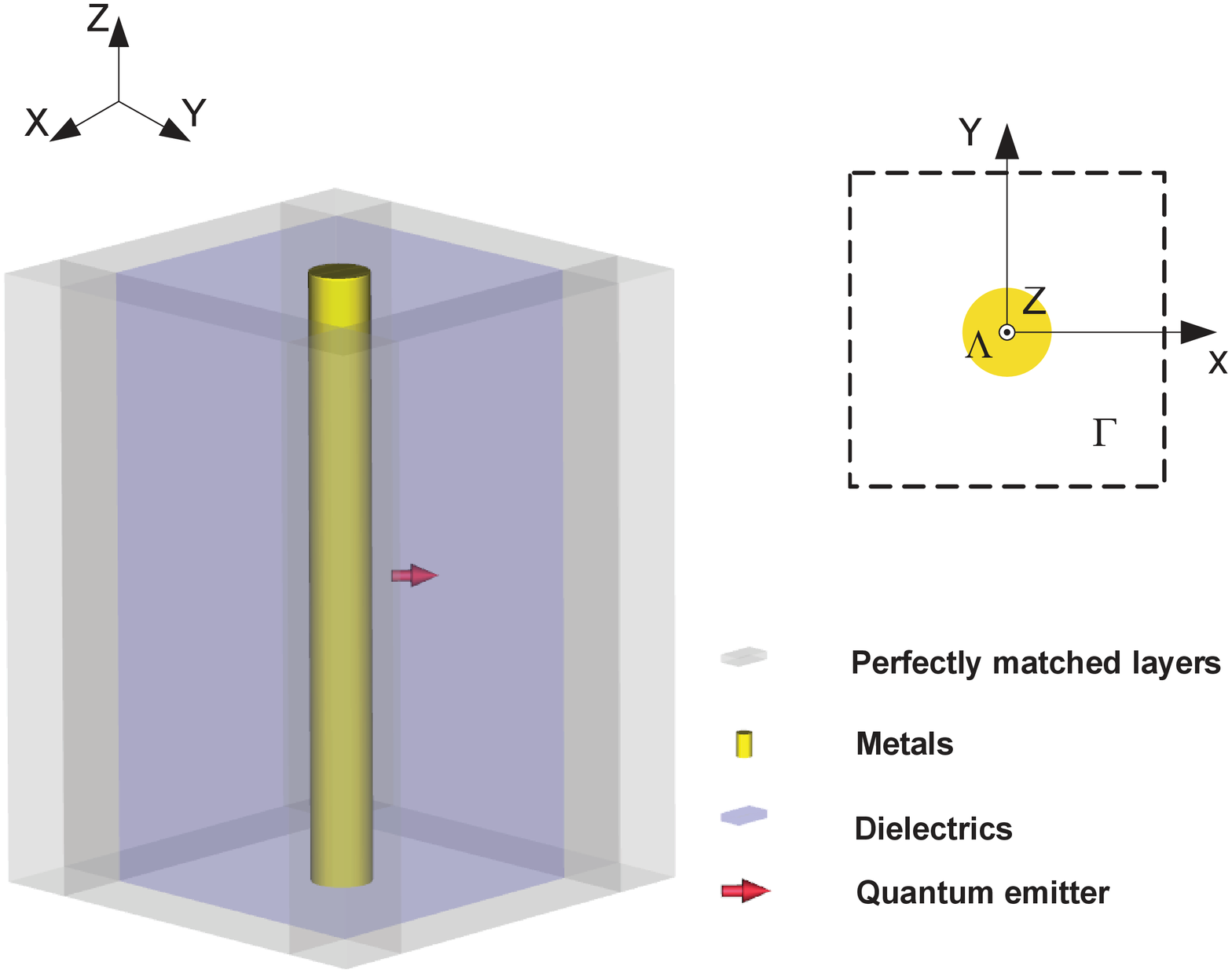}
\caption{\label{3D_model} A single quantum emitter coupled to a metallic nanowire. The grey transparent region represents the perfectly matched layers, the mode matching boundary condition is applied on the top and the bottom of the structure. The quantum emitter is implemented by an electric  line current.}
\end{figure}

It is crucial to truncate the computational domain properly. As shown in Fig.~\ref{3D_model}, we have two strategies to truncate the modeling domain: I) In the X-Y plane, the computation domain is truncated by the perfectly matched layers with thickness of half a wavelength in  vacuum. II) Along the Z-axis, the computation domain is terminated by mode matching boundary conditions, which will induce a certain amount of reflection from the radiation mode and the higher order plasmonic modes if they exist. Essentially, the mode matching boundary is an absorbing wall, which behaves as a sink of electromagnetic waves. There are different options of realizing the mode matching boundary to absorb a single mode, depending on whether the absorbed mode is TE, TM or a hybrid mode. For a pure TM or TE mode, it can be matched by simply applying the conditions,
\begin{subequations}\label{Pure_boundary}
\begin{align}
&\frac{1}{{\mu _r }}\bar n \times \nabla  \times \bar E(\bar r,\omega ) = \frac{{k_0^2 \varepsilon _r \bar E_t (\bar r,\omega )}}{{j\beta }}, & TM;\label{TM_boundary} \\
&\frac{1}{{\mu _r }}\bar n \times \nabla  \times \bar E(\bar r,\omega ) = j\beta \bar n \times \frac{1}{{\mu _r }}\bar n \times \bar E_t (\bar r,\omega), & TE;  \label{TE_boundary}
\end{align}
\end{subequations}
on the boundary, where $\beta$ is the propagation constant, and $\bar E_t (\bar r,\omega )$ is the tangential components of the dependent variable $\bar E(\bar r,\omega ) $ on the boundaries in the numerical model.
The mode matching boundary condition for the hybrid mode can be implemented as
\begin{equation}\label{Boundary_conditionhybrid3D}
\frac{1} {{\mu _r }}\bar n \times \nabla  \times \bar E(\bar r,\omega ) =  - j\omega \mu _0 \bar n \times \bar H_0,
\end{equation}
where $\bar E(\bar r,\omega )$ is the dependent variable solved in the 3D model, and $\bar H_0$ denotes the matched mode that is applied. In our model,  $\bar H_0$ corresponds to the fundamental hybrid mode supported by the plasmonic waveguide. It is  calculated from the 2D  eigenvalue problem, and is given by
\begin{equation}\label{Boundary_conditionhybrid}
\bar H_0 = \sqrt {\frac{{\gamma _{pl} P_0 }} {{P_{2d} }}} \bar H_{2d} e^{ - j\beta L_0 } =(H_{0x},H_{0
y},H_{0z}).
\end{equation}
Here, $P_{2d}$, $\bar H_{2d}$ and $\beta$ are the time averaged power flow, the magnetic field, and the propagation constant, respectively calculated from the 2D model, while $P_0$ denotes the normalization factor of the power emission in the 3D model, and $L_0$ represents the half length of the 3D model. Due to the losses of the metals, the magnitude of the magnetic field is a complex number. In order to guarantee that the phase of $E_x$ at the position of the emitter is zero when the emitter is oriented  horizontally, the extra phase $\phi=\arctan (\frac{{imag(E_{2D}^x )}}{{real(E_{2D}^x )}} )$ needs to be compensated, i.e., $\bar H_0 = \sqrt {\frac{{\gamma _{pl} P_0 }} {{P_{2d} }}} \bar H_{2d} e^{ - j(\beta L_0+ \phi)}$. In the 2D eigenvalue calculations, there are 6 components involved for the hybrid fundamental model,
the relations of which are tabulated in Table \ref{Efield_Hfield_relationship}. The  magnitudes of the magnetic field, $\left(H_{2d,m}^{x}, H_{2d,m}^{y}, H_{2d,m}^{z}\right)$, are the dependent variables, which are calculated directly from the 2D numerical model.

\begin{table}
\caption{\label{Efield_Hfield_relationship} The relation of the 6 field components for the fundamental hybrid mode}
\begin{tabular}{l|l}
\hline
Description & Relation   \\
     \hline
\hline
 \multirow{2}{3.8cm}{Tangential electric field, $s \in \left[ {x,y} \right]$}  & $E_{2D,t}^s =\frac{\beta }{{\omega \varepsilon }}(\bar n \times H_{2D,t} )_s  + \frac{j}{{\omega \varepsilon }}(\nabla _t  \times \bar nH_{2D}^n )_s $ \\
&\\
\hline
 Normal electric field & $E_{2D}^n =\frac{j}{{\omega \varepsilon }}\bar n \cdot (\nabla _t  \times H_{2D,t} ) $     \\
\hline
 \multirow{2}{3.8cm}{Tangential magnetic field, $s \in \left[ {x,y} \right]$}  & $H_{2D,t}^s =H_{2d,m}^{s}$ \\
&\\
\hline
Normal magnetic field &  $H_{2D}^n =j H_{2d,m}^{z}$ \\
\hline
\hline
\end{tabular}
\end{table}

\begin{figure}
\includegraphics[scale=0.75]{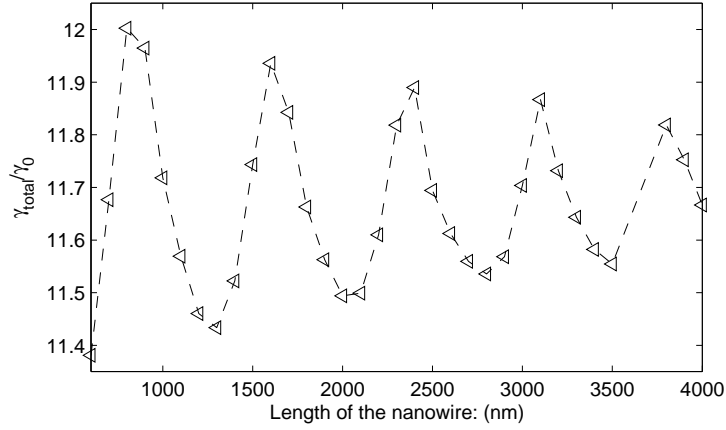}
\caption{\label{LengthdepedenceTMgodlwire} Length dependence study of the total decay rate for the metallic nanowire. The radius of the metallic nanowire is 20 nm, the distance of emitter to the wire edge is 30nm.}
\end{figure}

\begin{figure}
\includegraphics[scale=0.60]{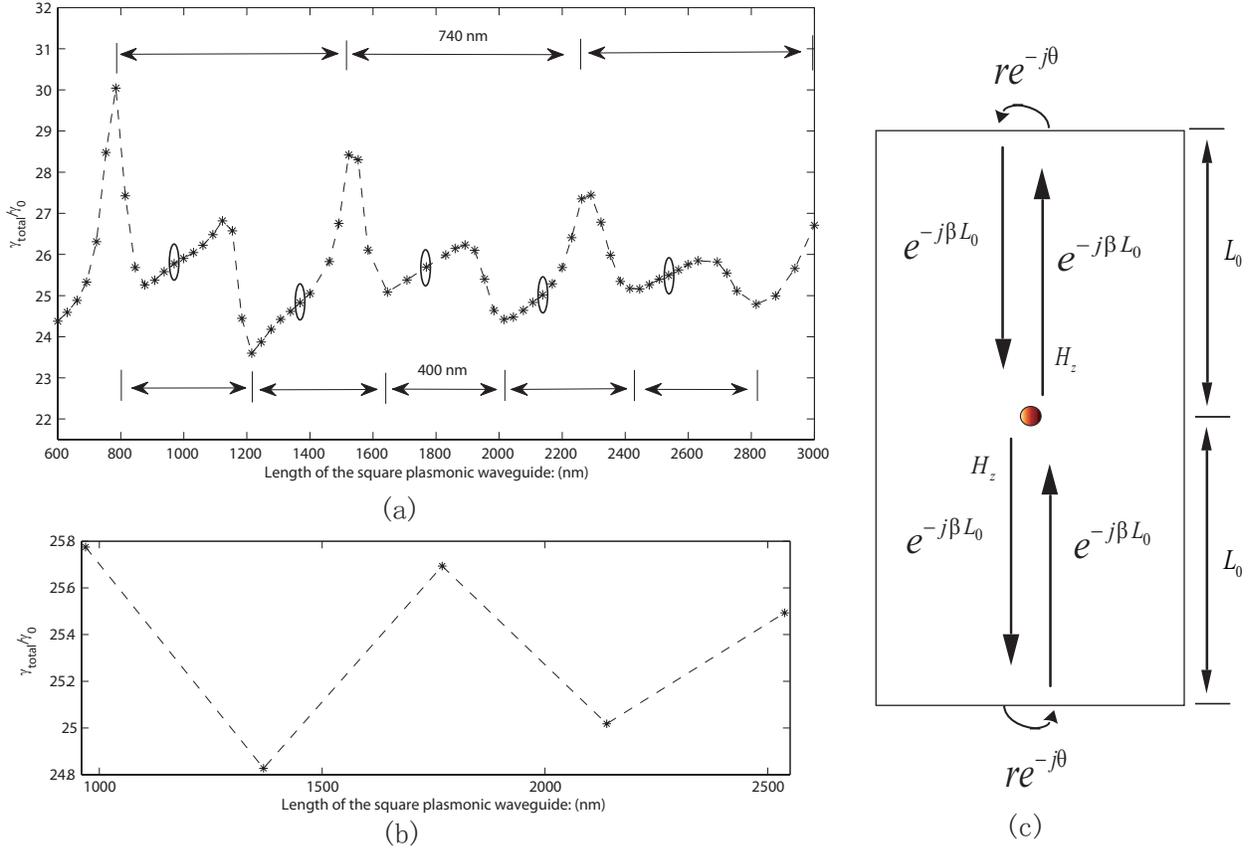}
\caption{\label{LengthdepedenceSquaregodlstrip} (a) Length dependence of the total decay rate for the square plasmonic waveguide. The side length is 30nm, the distance of the emitter to the edge of the square metal core is 20 nm. (a) Length dependence study with damped oscillations. (b) Length dependence  for the points from (a) (marked ellipses) where where $real(e^{ - j(\beta L_0+ \phi)})=0$ holds approximately. (c) Illustration of the reflection of  the normal   magnetic field of the fundamental hybrid in the 3D model, $r$ and $\theta$ are the reflection coefficient and phase shift respectively.}
\end{figure}

The total  decay rate, $\gamma_{total}$, is extracted from the total power dissipation of the current source coupled to the nearby metallic waveguide, $\gamma_{total}/\gamma_0= P_{total}/ P_0$, where $ P_{total}=1/2\int_V Re{(J^* \cdot E_{total})} dV$ is  the power dissipation of the current source coupled to the metallic waveguide, and $ P_{0}=1/2\int_V Re{(J^* \cdot E_{0})} dV$ is the emitted power by the same current source in vacuum. $P_0$ is a normalization factor, which is also used to normalize the power flow on the boundaries in  Eq.~(\ref{Boundary_conditionhybrid}). As demonstrated in Fig.~\ref{3D_model}, the field is generated by the current source, namely, the dipole emitter, which is implemented by a small electric line current. In our model, the dipole is oriented horizontally. For an  electric current source with finite size of $l$ ($l \ll \lambda_0$), and linear distribution of current $I_0$, the dipole moment of the source \cite{jackson} is, $\mu=j I_0 l/ \omega$. In order to avoid higher order multipole moments, the size of the current source should be restricted below a certain value. Our numerical test shows that the variation of the total power dissipation from the size dependence of the emitter is negligible  when the  size of emitter is shorter than 2 $nm$.

In order to check the validity of the mode matching boundary condition  we studied the length dependence of the total decay rate for  two different plasmonic waveguides. The length dependence of the total  decay rate $\gamma_{total}$ for the metallic nanowire  is shown in Fig.~\ref{LengthdepedenceTMgodlwire}. The fundamental mode supported by the metallic nanowire is TM, hence the mode matching boundary condition defined by Eq.~(\ref{TM_boundary}) is implemented. As can be seen from Fig.~\ref{LengthdepedenceTMgodlwire}, the variations in the total decay rate are reduced by increasing $L_0$, and the damped  oscillation of the total decay rate with $L_0$ indicates a certain amount of reflection from radiation modes, which is confirmed by the period of the oscillation ( equal to the wavelength in the media with $\varepsilon = 2$). We also see that the variation of the total decay rate due to the length dependence is below $\pm2.5\%$  due to the dominating excitation of the plasmonic mode  when $L_0$ is larger than 1 $\mu m$. Basically, the accuracy of $\gamma_{total}/\gamma_0$ relies on the length of the plasmonic waveguide, accordingly we estimate the relative error on the computed data is $\pm2.5\%$ in the following calculations for the metallic nanowire.

Regarding the square plasmonic waveguide, the  condition defined by Eq.~(\ref{Boundary_conditionhybrid}) is applied on the boundary to absorb the hybrid mode supported by the waveguide, where $\bar H_0$ is the  magnetic field for the matched field. As shown in Fig.~\ref{LengthdepedenceSquaregodlstrip}, there is also a damped oscillation of the total decay rate with the length of the computation domain, and the tendency of achieving higher accuracy  for $\gamma_{total}$ when  $L_0$ is lengthened, which is similar to the length dependence study of the total decay rate for the nanowire. Nevertheless, there are two  distinctions between the two plots: I) The variation of the total decay rate for the square plasmonic waveguide is much larger than that for the metallic nanowire; II) The variation of the total decay rate for the square plasmonic waveguide with $L_0$ primarily stems from the reflection of two different modes, which are indicated by two different periods in the damped oscillation. The reflection of the fundamental mode, which is supposed to be absorbed at the boundaries, is responsible for the oscillation with the period of 200nm, the other oscillation with the period of 370 nm results from the reflection of a quasi guided mode, denoted by $E_{qg}$.
The explanation is the following, the boundary condition defined by Eq.~(\ref{TM_boundary}) can completely absorb the matched pure TM mode, while it is not true for the boundary condition defined by Eq.~(\ref{Boundary_conditionhybrid}) for the hybrid mode, and a significant reflection from a quasi guided mode also exists for the square plasmonic waveguide. For the hybrid mode the last term in Eq.~(\ref{weak_form}) relies  not only on the tangential components of the electric (magnetic) field, but also on the normal component of the electric  (magnetic) field, which is intrinsically lost on the boundary in the vector element formulation of the 3D numerical model \cite{Jin}. Our interpretation is that, even though the normal component of the electric field can be included on the boundaries by Eq.~(\ref{Boundary_conditionhybrid}), the normal component of the magnetic field is essentially missing in the 3D numerical model with the square plasmonic waveguide, resulting in the reflections in our vector element formulation \cite{Kanellopoulos1995_normal_component, Webb_1993}. However, in Fig.~\ref{LengthdepedenceSquaregodlstrip}(a), it appears that the points for which  $real(e^{ - j(\beta L_0+ \phi)})=0$ holds approximately converge quickly with minimum impact of the reflection from the fundamental hybrid mode. The mode $E_{qg}$, with effective wavelength of $740.07$ nm, is characterized by the material properties of the waveguide  and is rather insensitive to the size of the metallic core. Compared with other quasi guided modes or radiation modes, the mode $E_{qg}$ has a relatively significant contribution to the $\gamma_{total}$, the normalized spontaneous emission rate  is 0.107. Since no extra effort is made to prevent the reflections from any components of  the mode $E_{qg}$, it is understandable that the induced reflections give rise to several  peaks in Fig.~\ref{LengthdepedenceSquaregodlstrip}(a).

The normal component of the magnetic field of the fundamental mode in the 3D model can be obtained by a 2D eigenvalue calculation, $H_{n,l}= \sqrt {\frac{{\gamma _{pl} P_0 }} {{P_{2d} }}} H_{2d} ^n e^{ - j(\beta l+ \phi)}$, where $l$ is the distance from the observation plane to the emitter. Similarly, the reflected  normal component of the magnetic field at the position of the emitter can be obtained by taking into account the phase shift due to propagation and reflection, $H_{n,0} ^r= r\sqrt {\frac{{\gamma _{pl} P_0 }} {{P_{2d} }}} H_{2d} ^n e^{ - j(2\beta L_0+ \phi+\theta)}$, as shown in Fig.~\ref{LengthdepedenceSquaregodlstrip} (c). The reflected normal component of the magnetic field will ``generate'' a perturbation term $E_x^r$ to the original $E_x$ component, the real part of which is integrated to calculate the total power dissipation. According to  Table \ref{Efield_Hfield_relationship}, the reflected term $E_x^r$ from the fundamental hybrid mode is given by
\begin{equation}\label{Ex_reflection}
E_x^r=-\frac{1}{{\omega \varepsilon }}(\nabla _t  \times \bar n(r\sqrt {\frac{{\gamma _{pl} P_0 }} {{P_{2d} }}}  H_{2d,m}^{z} e^{ - j(2\beta L_0+ \phi+\theta)} ))_x.
\end{equation}
The real part of $E_x^r$ can be zero when $L_0$ is appropriately chosen, therefore, the obtained total decay rates are expected to approach the true value more closely due to the vanishing contribution of $E_x^r$ to the total decay rate. In Fig.~\ref{LengthdepedenceSquaregodlstrip}(a), at the points  with marked ellipses, the half model length $L_0$ fits the requirement ($real(E_x^r)=0$), and we also found that the phase shift $\theta$ is required approximately to be $\pi/2$. Further examining the phase shift $\theta$ involves technical details regarding the implementation of the vector element formulation of the finite element method, which is beyond the scope of the present paper, and we refer to the references \cite{Jin, Kanellopoulos1995_normal_component, Botha_2006, Webb_1993}. From Fig.~\ref{LengthdepedenceSquaregodlstrip}(b), we estimate the relative error on the computed data for the square plasmonic waveguide to be $\pm2\%$, when $L_0$ is larger than 1 $\mu m$.

\section{Result and discussion}
\begin{widetext}
\begin{figure}
\includegraphics[scale=0.6]{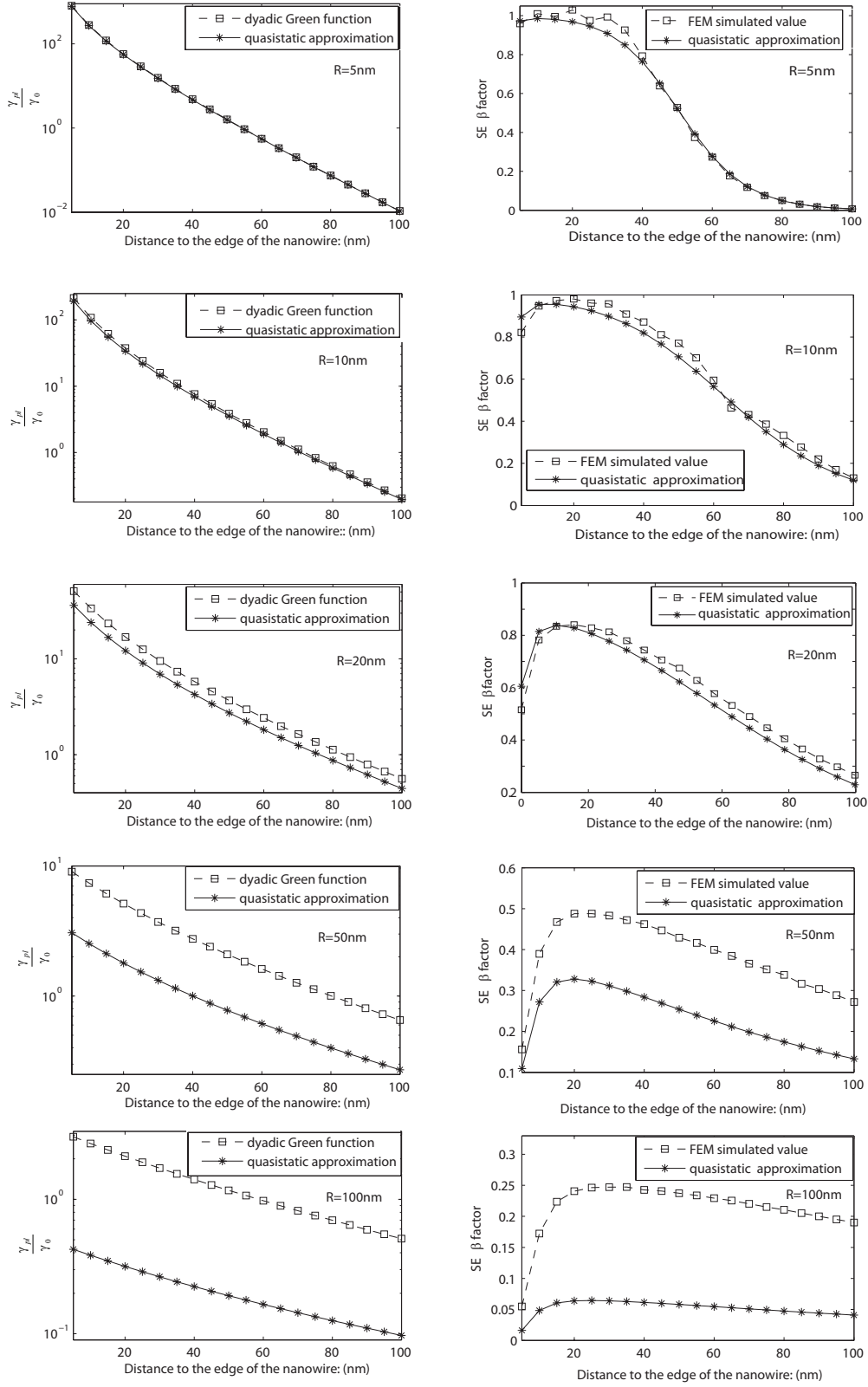}
\caption{\label{compare_FEM_quasi} Comparison of FEM simulated results based on the dyadic Green function  with the quasistatic approximation for the metallic nanowire.}
\end{figure}
\end{widetext}

\begin{figure}
\includegraphics[scale=0.75]{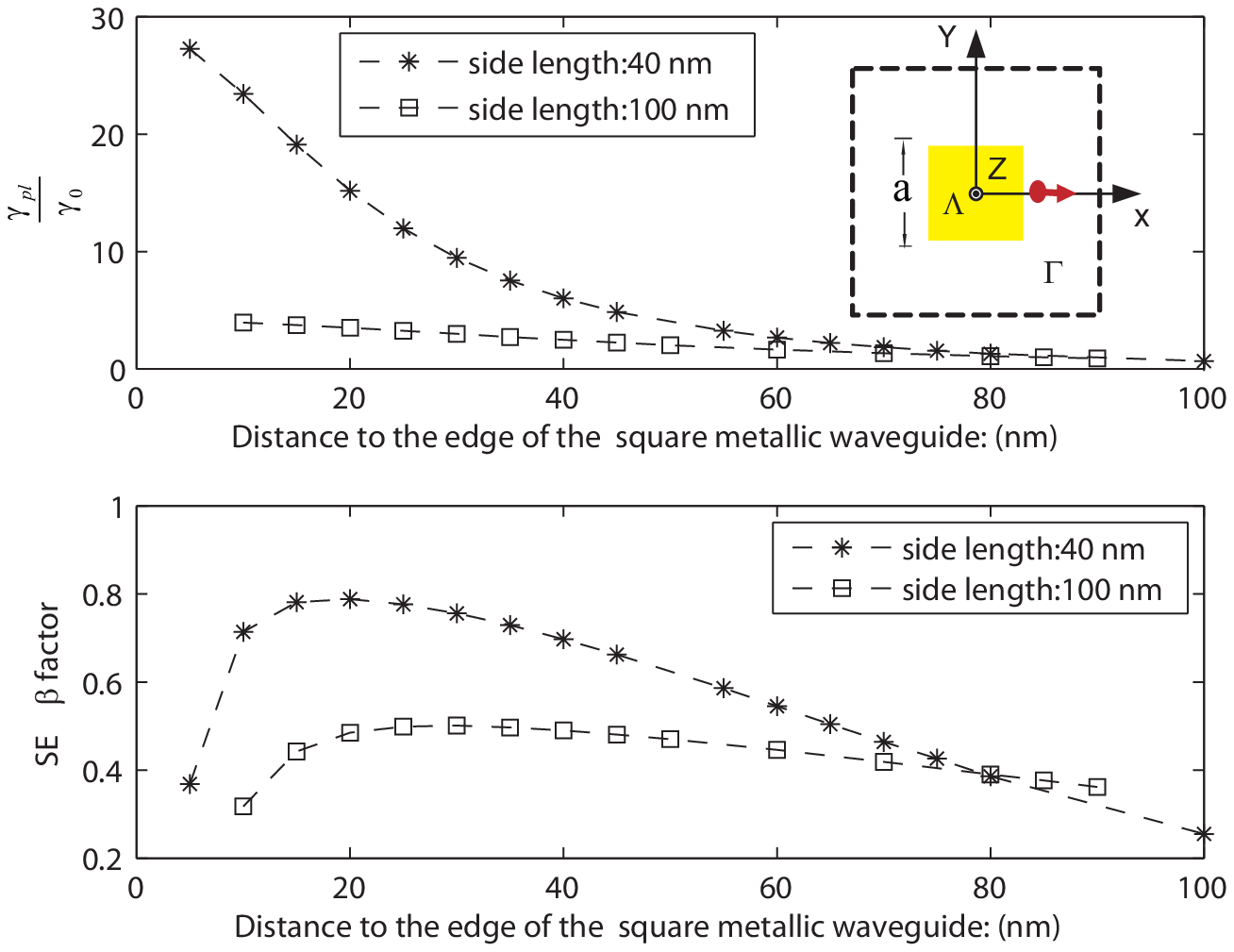}
\caption{\label{square_plasmonic_waveguide} Distance dependence of the plasmonic decay rates and spontaneous emission $\beta$ factors for the square plasmonic waveguide.}
\end{figure}
In this section,  the numerical method is applied to the two cases, the metallic nanowire  and the square plasmonic waveguide. For the metallic nanowire, we compared our numerical calculations  with the quasistatic approximation, which was studied by Chang  et al \cite{chang035420}. As can be seen in Fig.~\ref{compare_FEM_quasi}, our numerical results agree well with the quastatic approximation when the radius of the metallic nanowire is less than 20 nm, while it is  5-10 times larger when the radius is 100 nm. The deviation between the two methods increases when the radius becomes larger. The results of the comparison can be understood if one realizes that the quasistatic approximation is valid only when the radius is much smaller than the wavelength.  For the large wires, which has also been pointed out by Chang  et al. \cite{chang035420}, the full electrodynamic solutions predict significantly larger values of $\gamma _{pl} /{\gamma _0 }$ and the  $\beta$ factor. The quasistatic approximation assumes that the magnetic field vanishes, thus the obtained solution of the electric field simply behaves as a static field. It is the same for the plasmonic mode, the penetration length of which in the dielectric medium  is considerable shorter than that obtained  from the full electrodynamic solutions, and  therefore the quasistatic calculation predicts a lower coupling efficiency. In summary, our numerical calculation is consistent with the quasistatic approximation for the nanowire radii approaching 0, and the values from finite element simulation are generally larger than those obtained from the quasistatic approximation when the radius is beyond 20 nm.

We also studied the coupling of the quantum emitter with the square plasmonic waveguide. As shown in the inset in  Fig.~\ref{square_plasmonic_waveguide}, the quantum emitter is oriented along the $X$ axis, and the distance dependence of the plasmonic decay rates and spontaneous emission $\beta$ factors is calculated as function of emitter position along the $X$ axis. With optimized side length of the waveguide and distance of the emitter to the edge of the waveguide, the $\beta$ factor can  reach $80\%$.

\section{Conclusion}

In conclusion, we developed a self-consistent model to study the spontaneous emission of a quantum emitter at nanoscale proximity to a plasmonic waveguide using the finite element method. The dyadic Green function of the guided modes supported by the plasmonic waveguide can be constructed numerically from the eigenmode analysis, and subsequently the normalized decay rate into the plasmonic channel can be extracted. The 3D finite element model is also implemented to calculate the total decay rate, including the radiative decay rate, nonradiative decay rate, and the plasmonic decay rate. In the 3D model, it is assumed that only one guided plasmonic mode is dominatingly excited, which is normally true when the size of the cross section of the plasmonic waveguide is below 100 nm. Under such condition, the spontaneous emission $\beta$ factor is calculated. We compared our numerical approach with the quasistatic approximation for the gold nanowire. We observe agreement with the quasistatic approximation for radii below 20 nm, where the quasistatic approximation is valid. For larger radii the FE simulation predicts approximately 5 times larger values. This is reasonable since the numerical model takes into account wave propagation, whereas the quasistatic approximation calculates the static field. We also applied our numerical model to calculate the spontaneous emission of a quantum emitter coupled to a square plasmonic waveguide. The numerical calculations shows that  spontaneous emission $\beta$ factor up to $80\%$ can be achieved for a horizontal dipole emitter, when the distance and the side length are optimized.

\section{Acknowledgment}
The authors would like to thank Anders S. S\o rensen, Thomas S\o ndergaard, Andrei Lavrinenko and Darrick Chang for fruitful discussions. We gratefully acknowledge support from  Villum Kann Rasmussen Fonden via the NATEC center.

\bibliography{yuntianbib}
\end{document}